\documentclass[12pt,aps,nofootinbib]{revtex4}
\usepackage{amsmath}
\usepackage{graphicx}

\renewcommand{\vec}[1]{\boldsymbol #1}

\newcommand{\dtn}[1]{\frac{d^3{\vec #1}}{(2\pi)^3}\,}
\begin{document}

\title{ Nonperturbative Renormalisation Group Flow for ultracold Fermi-gases in unitary limit.}
\author{Boris Krippa$^{1}$}
\affiliation{$^1$ School of Physics and Astronomy,
The University of Manchester, M13 9PL, UK}
\date{\today}
\begin{abstract}
We study the exact renormalisation group flow for ultracold Fermi-gases in unitary regime. 
 We introduce a pairing field  
 to describe the formation of the Cooper pairs, and take a simple ansatz for 
the effective action. Set of approximate flow equations for 
the effective couplings including boson and fermionic fluctuations is derived. 
 At some  value of the running scale, the system undergoes  a phase transition to a gapped 
phase. The values of the energy density, chemical potential, pairing gap  and the corresponding proportionality constants
relating the interacting and non-interacting Fermi gases are calculated. Standard mean field results are recovered 
if we omit the boson loops.

\end{abstract}
\maketitle

The physics of ultracold fermionic gases has recently drawn much attention as it provides an exciting possibility of studying the regime where 
the dynamics of the many-body system becomes independent of the microscopical details of the underlying interaction between two fermions.
This regime can be probed using  the technique of Feshbach resonances \cite{Fe}  when the scattering length is
 tuned to be much larger then the average inter-particle separation.
The idealised case of the infinite scattering length $a$ is often referred to as the unitary regime (UR).
In the $a\rightarrow - \infty$ limit the ground state energy per particle is proportional to that of the non-interacting Fermi gas.

\begin{equation}
E_{GS}=\xi E_{FG}=\xi \frac{3}{5}\frac{k^{2}_{F}}{2 M},
\end{equation}
where $M$ and $k_F$ are the fermion  mass and Fermi momentum correspondingly and   $\xi$ is the universal
 proportionality constant, which does not depend on the details of the interaction or type of fermions.
 The other dimensional characteristics of the cold Fermi-gas in the UR such as paring energy $\Delta$ or chemical potential $\mu$
can also be represented in the same  way

\begin{equation}
\mu = \eta E_{FG} , \qquad \Delta = \epsilon E_{FG}
\end{equation}
 The infinite scattering length implies nonperturbative treatment. The most ``direct''
nonperturbative method is based on the lattice field theory \cite{Ca, As}. However, being potentially the most powerful approach,
 lattice simulations still have many limitations related to finite size effects, discretization errors etc which may even become amplified
in certain physical situations (the system of several fermion species is one possible example). All that makes the development of the analytic 
approaches indispensable. Several such approaches have been suggested so far. The (incomplete) list includes the 
Effective Field Theory (EFT) motivated 
formalism exploring the systematic expansion in terms of dimensionality of space \cite{So} and somewhat similar approach, 
based on $1/N$ expansion \cite{Sa}. More phenomenological approaches using the density functional method and 
many body variational formalism have been developed in \cite{Bu} and \cite{Ha}.

In this paper we consider the cold Fermi gas in UR in the framework of the Exact Renormalisation Group (ERG) method suggested in \cite{We}
and applied to the nonrelativistic many-fermion system with pairing in \cite{Kr}.
Different aspects of the  ERG approach  to nonrelativistic systems have also  been extensively studied in several subsequent  papers \cite{Kr1,Kr2,Sal,Di}.
 Being spiritually related 
to the EFT based approaches the ERG formalism is however fully nonperturbative  and makes  use of EFT as a guide to choose the ansatz for the effective action
and to fix  boundary conditions. The technical details of the approach in the context of the nonrelativistic many-fermion systems
  were described  in \cite{Kr1, Kr2} so that here we give only a short account of the formalism.

The central object of the ERG formalism is the average effective action 
$\Gamma_k$ which coincides with the bare action at the beginning of the evolution when the scale
 $k = \Lambda$ (with $\Lambda$ being a starting scale) and is a full quantum action when $k = 0$. The average effective action 
satisfies the following flow equation 

\begin{equation}
\partial_k\Gamma=-\frac{i}{2}\,{\rm Tr}\left[(\partial_kR)\,
(\Gamma^{(2)}-R)^{-1}\right],
\end{equation}
where
\begin{equation}
\Gamma^{(2)}=\frac{\delta^2\Gamma}{\delta\phi_c\delta\phi_c},
\end{equation}
and $R_k$ is the regulator satisfying the following two conditions $R_{k\rightarrow 0}\rightarrow 0$ and 
$R_{k\rightarrow \infty}\rightarrow k^2$. To find a (approximate) solution of the ERG flow equations 
one needs to choose the ansatz for the effective action and fix the corresponding boundary conditions represented by the bare action.
There are no strict quantitative criteria for choosing the ansatz so that it seems reasonable to work with the 
 ``relevant'' degrees of freedom and include the interaction terms satisfying (and allowed by) all possible symmetry 
constraints. In our case the degrees of freedom are the strongly interacting fermions in the limit of 
 the infinite scattering length. At the starting scale the medium parameters like Fermi-momenta, chemical potentials etc
play a little role so that at this scale the average effective action is just the bare action with the standard
 four-fermion interaction in vacuum. One possible choice for the bare action is the simplest attractive 
EFT-motivated four-fermion pointlike interaction with the lagrangian

\begin{equation}
{\cal L}_i=-\frac{1}{4}\,C_0\left(\psi^\dagger\sigma_2\psi^{\dagger{\rm T}}\right)
\left(\psi^{\rm T}\sigma_2\psi\right).
\end{equation}

With a decreasing scale the role of the many-body effects becomes more and more important
and at some scale one may anticipate the Cooper instabilities, symmetry breaking, formation of 
the correlated fermion pairs etc to occur. Taking all that into account we write the ansatz for the effective 
action $\Gamma_k$ in the following form
\begin{eqnarray}
\Gamma[\psi,\psi^\dagger,\phi,\phi^\dagger]&=&\int d^4x\,
\left[\phi^\dagger(x)\left(Z_\phi\, i \partial_t 
+\frac{Z_m}{2m}\,\nabla^2\right)\phi(x)-U(\phi,\phi^\dagger)\nonumber\right.\\
&&\qquad\qquad+\psi^\dagger\left( Z_\psi (i \partial_t+\mu)
+\frac{Z_M}{2M}\,\nabla^2\right)\psi\nonumber\\
&&\qquad\qquad\left.- g \left(\frac{i}{2}\,\psi^{\rm T}\sigma_2\psi\phi^\dagger
-\frac{i}{2}\,\psi^\dagger\sigma_2\psi^{\dagger{\rm T}}\phi\right)\right],
\end{eqnarray}
We introduce the effective pairing field $\phi$ and used the standard Hubbard-Stratonovich transformation to cancel
the four-fermion interaction. The couplings $Z_{(m,M,\phi,\psi)}$ and $g$ all run with the scale. We also include 
the kinetic term for the pairing field needed to compute the boson loop contributions. Note that our definition of the effective 
potential includes the term $- 2\mu Z_{\phi}\phi^\dagger\phi$ which describes the coupling of the pairing field to the chemical potential.
We expand the effective potential about its minimum

\begin{equation}
U(\phi,\phi^\dagger)= u_0+ u_1(\phi^\dagger\phi-\rho_0)
+\frac{1}{2}\, u_2(\phi^\dagger\phi-\rho_0)^2,
\end{equation}
where the $u_n$ are defined at the minimum of the potential,
$\phi^\dagger\phi=\rho_0$. The coefficients at the quadratic (in fields) term determine the phase of the system.
When $u_1 >$  0 the system is in the symmetric phase with a trivial vacuum $\rho_0=0$. At some critical scale the coefficient
$u_1$ approaches zero and the system undergoes the transition to the broken (or condensed) phase with $\rho_0\neq 0$ so that in this phase

\begin{equation}
U(\phi,\phi^\dagger)= u_0+\frac{1}{2}\, u_2
(\phi^\dagger\phi-\rho_0)^2 + .....,
\end{equation}
The renormalisation factors can also be expanded about $\rho = \rho_0$
\begin{equation}
Z_{\phi}(\phi,\phi^\dagger)= z_{\phi 0} + z_{\phi 1}(\phi^\dagger\phi-\rho_0) + .....,
\end{equation}
The other renormalisation factors can be expanded in the same way.\\
In this paper we take into account the terms in the expansion of the effective potential up to quartic order
in the fields. The minimum $\rho_0$ of the effective potential evolves with the scale 
in the condensed phase and all the coefficients of  the expansion should depend 
on both the scale and $\rho_0$. There are few options of how to organise the evolution 
of the system. The choice would  depend on the physical quantities which 
one would like to get  at the end of the evolution. The obvious and the most general
one is just to run all quantities of interest. The other (simpler) way to extract essentially the same information 
is to fix some parameters like chemical potential or particle number density  and run the rest.
Fixed particle number density and evolving  chemical potential
seems more interesting as it gives the potential possibility of going to the BEC regime 
where the chemical potential eventually becomes negative. In this case the coefficients 
$u_n$ and the renormalisation factors will depend on the running scale both explicitly
and implicitly via the dependencies on  $\rho_{0}(k) $ and $\mu(k)$.
We can, therefore, define a total derivative

\begin{equation}
\frac{d}{dk}=\partial_k+\frac{d\rho_0}{dk}\,\frac{\partial}{\partial\rho_0}
+\frac{d\mu}{dk}\,\frac{\partial}{\partial\mu}.
\end{equation}
Applying this to $n = -\partial U/\partial \mu$ at $\rho=\rho_0$ and assuming the
 constant particle number density we get
\begin{equation}
-2z_{\phi 0}\,\frac{d\rho_0}{dk}+\chi\,\frac{d\mu}{dk}
=-\left.\frac{\partial}{\partial \mu}
\Bigl(\partial_k  U\Bigr)\right|_{\rho=\rho_0}.
\label{eq:muevol}
\end{equation}
where we defined 
\begin{equation}
\chi=\left.\frac{\partial^2 U}{\partial \mu^2}
\right|_{\rho=\rho_0}.
\end{equation}
The other ERG flow equations can be obtained in a similar way so that after some algebra we get

\begin{eqnarray}
\frac{du_0}{dk}+n\,\frac{d\mu}{dk}
&=&\left.\partial_k U\right|_{\rho=\rho_0},\\
\noalign{\vskip 5pt}
-u_2\,\frac{d\rho_0}{dk}+2z_{\phi 0}\,\frac{d\mu}{dk}
&=&\left.\frac{\partial}{\partial \rho}
\Bigl(\partial_k U\Bigr)\right|_{\rho=\rho_0},\\
\noalign{\vskip 5pt}
\frac{du_2}{dk}-u_3\,\frac{d\rho_0}{dk}+2z_{\phi 1}\,\frac{d\mu}{dk}
&=&\left.\frac{\partial^2}{\partial \rho^2}
\Bigl(\partial_k U\Bigr)\right|_{\rho=\rho_0},\\
\noalign{\vskip 5pt}
\frac{dz_{\phi 0}}{dk}-z_{\phi 1}\,\frac{d\rho_0}{dk}+\frac{1}{2}\,\chi'\,
\frac{d\mu}{dk}
&=&-\,\frac{1}{2}\left.\frac{\partial^2}{\partial \mu\partial\rho}
\Bigl(\partial_k U\Bigr)\right|_{\rho=\rho_0}.
\end{eqnarray}
where we have defined
\begin{equation}
\chi'=\left.\frac{\partial^3 U}{\partial \mu^2\partial\rho}
\right|_{\rho=\rho_0}
\end{equation}
Note that we introduce the coefficient $u_3$ which corresponds the higher order terms in the expansion of the effective potential. 
This coefficient occurs when we act on $\frac{\partial^2}{\partial \rho^2} U$ with the above defined  total derivative.

The set of evolution equations in the symmetric phase can easily be recovered 
using the fact that chemical potential does not run in symmetric phase and that
$\rho_0 =0$. All the higher order terms such as $\chi$,$\chi'$, $u_3$ and $z_{\phi 1}$
were calculated from the mean field type of expression when the boson loops are
neglected and the effective potential can be calculated explicitly (see below).
The functions $\rho_{0}(k)$ and $\mu(k)$ which determine the physical energy gap and chemical
potential in the limit $k \rightarrow 0$ were computed in \cite{Kr1}. In this paper we focus on
 the field independent part of the effective potential $u_0$ which is related to the energy density of the  Fermi-gas in the UR.
As can be seen from the evolution equations the coefficient  $u_0$ is not coupled to the rest
but its value depends on running chemical potential $\mu(k)$ so that to find $u_0$ we have to 
solve the whole system of the flow equations.

The explicit expressions for the driving term $\partial_{k} U$ can be obtained from the effective action 
after the straightforward algebra and has the following form
\begin{equation}
\partial_k U = -\int\frac{d^3{\vec q}}{(2\pi)^3}\,\frac{E_{FR}}
{\sqrt{E_{FR}^2+\Delta^2}}\partial_k R_F  + \frac{1}{2Z_\phi}\int\frac{d^3{\vec q}}{(2\pi)^3}\,
\frac{E_{BR}}{\sqrt{E_{BR}^2-V_B^2}}
\,\partial_kR_B.\label{eq:potevol}
\end{equation}
where
\begin{equation}
E_{BR}(q,k)=\frac{Z_m}{2m}\,q^2+u_1
+u_2(2\phi^\dagger\phi-\rho_0)+R_B(q,k), \qquad V_B= u_2\phi^\dagger\phi=u_2\rho
\end{equation}
and 
\begin{equation}
E_{FR}(q,k,\mu)=\frac{Z_M}{2 M}(q^2-p_F^2)+R_F(q,k,\mu).
\end{equation}
The other driving term can be obtained by taking the corresponding derivatives of $\partial_k U$.
The evolution equation as it is written above is still not enough to extract the energy density as it suffers
from the divergence in the UV limit when $k \rightarrow \Lambda$. Therefore, one needs to make a subtraction 
based on a physical assumption that the energy density must be a constant, equal to that of the free Fermi gas
 when $u_{1}=u_{2}=...=u_{n}=0$. The modified flow equation for the effective potential 
can be written as 

\begin{equation}
\partial_k  U = -\int\frac{d^3{\vec q}}{(2\pi)^3}\,(1 - \frac{E_{FR}}
{\sqrt{E_{FR}^2+\Delta^2}})\,\partial_kR_F + \frac{1}{2Z_\phi}\int\frac{d^3{\vec q}}{(2\pi)^3}\,
(1 - \frac{E_{BR}}{\sqrt{E_{BR}^2-V_B^2}})
\,\partial_kR_B.\label{eq:potevol1}
\end{equation}
Note that the flow does not change as the added term is field independent. 
We utilise the type  of the  cut-off functions suggested first in \cite{Li}(see further discussion in \cite{Li1})
 for the boson ERG flow    and  in \cite{Kr2}
for the fermionic case. This form of the  cut-off functions allows for the significant practical simplifications when calculating
the loop diagrams. One notes that the  cut off function for fermions can be written in different forms. However, all of them should contain the 
$sgn$ function reflecting the particle and/or hole ``faces'' of the in-medium fermion. In this paper we use the cut off function in the form,
considered in \cite{Di}   
\begin{equation}
R_F = \frac{1}{2M}\left[(k^{2} sgn(q - p_\mu)- (q^2 -  p^{2}_{\mu}))\right]\theta(k^2  - \mid q^2  - p^{2}_{\mu}\mid),
\end{equation}
where $p_\mu = (2 M \mu)^{1/2}$, and
\begin{equation}
R_B = \frac{1}{2m}(k^2 - q^2)\theta(k -q). 
\end{equation}

 The boson loops can be neglected at high scale and the flow equation 
can be integrated explicitly resulting in the following expression for the effective potential
in the mean field (MF) approximation

\begin{equation}
U^{\text{MF}}(\rho, \mu, k) = \int \dtn{q}
\left(E_{FR}(\rho, \mu, k) - \sqrt{E_{FR(\rho, \mu, k)}^2+g^2\rho}\right)+ C,\label{eq:UF1}
\end{equation}
where C is the constant of integration. After some algebra the expression for  $U^{\text{MF}}$ can be rewritten as

\begin{equation}
U^{\text{MF}}(\rho, \mu, k)=\int \dtn{q}
\left(E_{FR}(q, \mu, k) + \frac{g^2\rho}{2\epsilon_{q}}
-\sqrt{E_{FR}(q, \mu, k)^2+g^2\rho}\right) - \frac{M g^2\rho }{4\pi a},  
\end{equation}
where $\epsilon_{q} = E_{FR}(q, 0, 0)$.

Differentiating with respect to $\rho$ and setting the derivative and running scale equal to
zero, we find that $\Delta^2$ at the minimum satisfies
\begin{equation}
-\,\frac{M}{4\pi a}+\frac{1}{2}\int\frac{d^3{\vec q}}{(2\pi)^3}\,
\left[\frac{1}{E_{FR}(q,0,0)}-\frac{1}{\sqrt{E_{FR}(q,p_F,0)^2+\Delta^2}}
\right]=0.
\end{equation}
This is exactly the gap equation derived, for example in \cite{PB99}.

To get the number density of fermions, we can differentiate $ U(\rho,\mu,0)$
with respect to $\mu$. This gives 
\begin{equation}
n=\int\frac{d^3{\vec q}}{(2\pi)^3}\,\left[1
-\frac{E_{FR}(q,p_F,0)}{\sqrt{E_{FR}(q,p_F,0)^2+\Delta^2}}\right],
\end{equation}
again in agreement with Ref.~\cite{PB99}. One can therefore conclude that the standard MF result can be reproduced within 
NRG if the boson fluctuations are neglected.

The boundary conditions for the coefficients $u_i$ can be obtained by differentiating the expression for $U^{\text{MF}}$ with respect to 
$\rho$ at the starting scale $k = k_{st}$

It is worth mentioning that in this paper we include running of $\mu, \rho_0, u_{i}'s$ and the renormalisation factor $Z_{\phi}$. This is the minimum
set of running parameters needed to go beyond the mean field approximation. The other couplings are held fixed at their initial values. 

The useful quantity to check the consistency of the approach is the boson scattering length $a_{B}$. It is well known  that the  MF
calculations lead to the relation $a_{B} = 2 a_{F}$, where  $a_{F}$ is the fermion scattering length. Deviation from this result is due 
to the boson loop effects and therefore goes  beyond the  MF approximation. In our approach the boson scattering length is given by the 
relation  $a_{B} = 2 u_2 / Z^{2}_{\phi}$. Using the cut-off in the form specified in the Eqn. (22) and calculating the values of  $u_2$ and $Z^{2}_{\phi}$
in the MF approximation it can easily be demonstrated that the relation $a_{B} = 2 a_{F}$ is indeed satisfied.

The calculations with boson loops lead to the relation $a_{B} = 1.13 a_{F}$. It is still quite  
 far away from the relation $a_{B} = 0.6 a_{F}$ found in the full 4-body calculations of Ref. \cite{Pet}. It means
that the present truncation, while providing useful tool to go beyond the MF level, is still too crude to realistically describe
the effects of boson loops at least for the boson scattering length. We note in passing that neither Yukawa coupling nor
fermionic renormalisation constants run in vacuum.It is worth mentioning that  similar ERG studies \cite{Di} resulted in relation $a_{B} = 0.91 a_{F}$.
 The nature of the difference 
between two otherwise similar calculations is not clear at present. This issue clearly requires further investigations.

 The  boson loop contributions are found to be small in the unitary regime
 when medium effects are included as the boson rescattering effects 
will  be partially suppressed in medium. More definite conclusion can be drown after 
 a number of other effects, for example evolution of all the couplings,  is  taken into account. 

Note that  it is  important to keep boson loops for theoretical consistency as it leads to the convex effective potential. 
The point is that in the vicinity of the physical point $k = 0$ the coupling $u_2$ turns out to be vanishingly small so that the potential becomes flat with 
the minimum shifted from the origin. We emphasise  that the effective potential retains its ``mexican hat'' form at any finite scale.
With a decreasing scale the 
bump becomes less and less pronounced so  the effective potential eventually evolves into the convex form.  Without the boson loop
 contribution the $u_2$ coupling
is finite at $k \rightarrow 0$ and not small so that the  convexity property of the effective potential is  missing in agreement with the  known results.

\begin{figure}
\hspace*{\fill}\includegraphics[height=5.0cm]{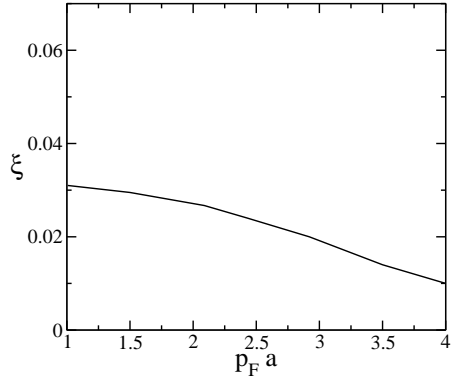}\hspace*{\fill}
\caption{Diviation of chemical potential from its MF value as a function of $p_{F}a$.}
\label{fig:dif1}
\end{figure}

In addition, smallness of the boson loop contributions in the unitary regime qualitatively agrees with the conclusion
 made in \cite{Str} where the effects of boson loops 
were included via the corresponding self-energy corrections.  According to Ref. \cite{Str} the boson loops contribution at zero temperature
is relatively small both in the unitary regime as well as in the BCS/BEC regions. We show in Fig.1 the deviation of the full chemical potential from 
its MF value as a function of $(p_F a)^{-1}$ in the BEC regime. This deviation is defined as $\xi = \mu (MF) - \mu (Full)$.
It is seen from Fig.1 this quantity is small compared to the typical value of chemical potential ($\mu \sim O(1)$ at $(p_F a)^{-1} \simeq 1$).
We emphasize that, although we believe that this conclusion is qualitatively correct, the actual value of the ``beyond-mean-field'' corrections
can be different if running of all the couplings is included. Besides, including higher order terms in the effective action can further change 
the contributions from the   ``beyond-mean-field'' effects. All that constitues the subject for the future studies. 
We stress that, despite of being clearly subleading order, the boson loops contribution to the physical observables
calculated in \cite{Str} is still larger then one obtained in this paper. To  understand the cause of this deviation one needs to establish one-to-one 
diagrammatic correspondence between NRG and the more traditional approach adopted in  \cite{Str}. It is  a difficult and still unsolved problem 
as ERG includes, in principle, an infinite (and probably mixed) set of diagrams so that one-to-one 
 correspondence between different contributions in the ERG approach and certain classes of diagrams may simply not exist.

Now we turn to the results. First observation is that the results become practically independent on the starting scale $\Lambda$
if $\Lambda\geq 10 p_F$. The phase transition to the condensed phase occurs at $k_{crit}\simeq p_F$. One notes that
 the calculations with the other types of cut-offs, both sharp and
smooth, lead to relatively close values of $k_{crit}$ \cite{Kr, Kr1}. The fact that all values of $k_{crit}$ are clustered around the value of $k_F$ makes
a good sense as at this scale  the system becomes sensible to the medium effects like Cooper instabilities, gap formation etc.

We found the values of $0.62, 0.96$ and $1.11$ for the universal coefficients $\xi, \eta$
and $\epsilon$ correspondingly.
The calculations without boson loops give $\xi_{MF} (\eta_{MF},\epsilon_{MF})  = 0.65 (0.98, 1.14)$. The uncertainties, related to the form of the regulator
are on the level $5 \%$.
The effect of the  boson loops is small. It holds for both optimised cutoff function used in this paper and for 
the smoothed theta function used in \cite{Kr}. One notes, however that  two cutoffs lead to different signs of the ``beyond-mean-field'' contribution.
In ideal case all the cutoffs should, of course, lead to the same results but in practise the unavoidable truncation of the effective action will always lead
to some uncertainties. Taking into account the smallness of the boson loop effects it is hard to see if this sign uncertainty is the result of 
truncation or just a numerical instabilities which are known to be larger for expressions involving step functions.

The obtained value for  $\xi$ is close  to
 the experimental data  from \cite{Ge},  $\xi = 0.74(7)$.
The other measurements give $\xi = 0.34(15)$ and $\eta = 0.6(15)$\cite{Bo}; $\xi = 0.32^{+0.13}_{-0.10}$ and $\eta = 0.53^{+0.13}_{-0.10}$ \cite{Ba}; 
$\xi = 0.46(5)$   and $\eta = 0.77(5)$\cite{Pa};
$\xi = 0.51(4)$ and $\eta = 0.85(4)$ \cite{Ki}. The epsilon expansion \cite{So} results in $\xi = 0.39$ and $\eta = 0.79$. Lattice simulations give
$\xi = 0.42(1),\eta = 0.71(1)$ and $\epsilon = 0.9(1)$ \cite{Ca}; $\xi = 0.22(3)$\cite{Lee}, $\xi = 0.44$ \cite{Bu1}, $\xi = 0.37$ \cite{Bu2},
 $\xi = 0.3$ \cite{Se};   $\xi = 0.41(2)$ and $\eta = 0.7$ \cite{As}.
Another ERG studies \cite{Di} give $\xi = 0.55$. As one can see 
both experiment and numerical simulations 
do not provide the coherent value of the $\xi$ constant so it is difficult to judge the quality of the numerical estimates 
provided by the  ERG calculations. One may only conclude that the ERG approach leads to the sensible values of the  universal coefficients
consistent with the experiment and lattice calculation but more detailed comparison  can be done when more accurate date are obtained.
We note, however that the value of  the universal parameters  are still somewhat higher then the ``world average''. One possible cause could 
be the neglection of the screening effects \cite{Go} which are known to decrease the values of the gap and energy density. 
Naive extrapolation of our results using the known value of the Gorkov - Melik-Barkhudarov's correction \cite{Go} indeed brings the values of the universal 
coefficients closer to the ``world average'' of the lattice and experimental data. Clearly,  this point requires  further 
analysis and the corresponding work is now in progress \cite{Kr3} (see also \cite{Di3}) . There are several other directions where the current ERG approach can further be 
developed. Firstly,as mentioned above, running of all coupling constants should be included. Secondly, the calculations with the entire effective potential 
taking into account the  screening effects should also
be performed \cite{Kr4}. Among the other physical applications one could mention the extension to  the finite temperature case and analysis of the
 deviations from the unitary limit.\\
{\bf Acknowledgements}\\
The author is grateful to  Mike Birse, Niels Walet, Judith McGovern and  Jean-Paul Blaizot  for the uncounted number of valuable discussions.

\end{document}